\documentclass[a4paper]{VisionStyle}
\usepackage{epsfig}

\begin{document}

\title{Morphology of Galaxy Clusters and Sunyaev-Zel'dovich effect}

\author{Ph.\,Jetzer\inst{1} \and P.\,Koch\inst{1} \and 
R.\,Piffaretti\inst{1}\inst{,2} \and D.\,Puy\inst{1}\inst{,2} \and 
S.\,Schindler\inst{3}\inst{,4}} 

\institute{
 Institute of Theoretical Physics, University of Z\"urich, 
Winterthurerstrasse, 190, 8057 Z\"urich, Switzerland
\and 
 Paul Scherrer Institute, Laboratory for Astrophysics, 5232 Villigen, 
Switzerland
\and 
Institute of Astrophysics, University of Innsbruck, 
Technikerstrasse 25, 6020 Innsbruck, Austria
\and
 Astrophysics Research Institute, Liverpool John Moores University, Twelve 
Quays House, Birkenhead CH41 1LD, UK}
\maketitle 

\begin{abstract}
Observations of the $X$-ray band wavelength reveal an evident 
ellipticity of many galaxy clusters atmospheres. The modeling of the 
intracluster gas with an ellipsoidal $\beta$-model leads to different 
estimates for the total gravitational mass and the gas mass fraction of the 
cluster than those one finds for a spherical $\beta$-model. An analysis 
of a recent {\it Chandra} image of the galaxy cluster RBS797 indicates a 
strong ellipticity and thus a pronounced aspherical geometry. A preliminary 
investigation which takes into account an ellipsoidal shape for this cluster 
gives different mass estimates than by assuming spherical symmetry. 
We have also investigated the influence of aspherical geometries of galaxy 
clusters, and of polytropic profiles of the temperature on the estimate 
of the Hubble constant through the Sunyaev-Zel'dovich (SZ) effect. We find 
that the non-inclusion of such effects can induce errors up to 40 \% on the 
Hubble constant value.

\end{abstract}

\section{Introduction}
Clusters of galaxies can be 
used to study how structures form on large scales. The formation and 
evolution of clusters depend very sensitively on cosmological parameters 
like the mean matter density $\Omega_m$ in the Universe. Thus it is of 
great importance to determine the dynamical state of clusters at different 
redshifts, see e.g. \cite{sabi2001}. 
\\
Geometry can give important insight in the 
dynamics of galaxy clusters. For example, the fitting of the cluster 
$X$-ray surface brightness with $\beta$-models usually provides different 
best fit parameters depending on the shape one assumes for the 
intracluster gas, but generally  the {\it classical} calculations of the mass 
of galaxy clusters suppose a spherical distribution of the density. 
\\
The SZ effect, \cite{zel1972}, offers the possibility to put 
important constraints on the cosmological models. Combining the temperature 
change in the cosmic microwave background due to the SZ effect and 
the $X$-ray emission observations, the angular distance to galaxy clusters 
and thus the Hubble constant $H_o$ can be derived. Nevertheless, 
geometrical shape of galaxy clusters can also introduce some 
{\it errors} on the analysis of the SZ effect. 
\\
In Section 2 we will analyze some consequences of the geometry 
of the clusters of galaxies and particularly on the mass, by taking as an 
example a recent observation of the galaxy cluster RBS797. Then we will 
describe, in Section 3, the influence of the cluster shape on the estimate of 
the SZ effect. In section 4 we will give an outlook on the role played by 
the geometry of galaxy clusters.
\section{Morphology of Galaxy Clusters}
The $\beta$-model, \cite{cava1976}, is widely used in $X$-ray astronomy 
to parametrise the gas density profile in clusters of galaxies by fitting 
their surface brightness profile. In this fitting procedure spherical symmetry
 is usually assumed, also in cases where the ellipticity of the surface 
brightness isophotes is manifest. For example \cite{Fab1984} 
showed a pronounced ellipticity of the 
surface brightness for the cluster Abell 2256, \cite{Al1993}
 obtained the same result for the profile of Abell 478 and 
\cite{Neu1997} for CL0016+16.\\
The asphericity of the observed surface brightness let us also ponder on the 
possible asphericity of the intracluster medium, 
which can be modelled with an ellipsoidal 
$\beta$-model rather than with the less accurate spherical one.\\
\cite{Hu1998} fitted the surface brightness of 
CL0016+16, which shows an axis ratio of major to minor axis of 1.176, with 
both circular and elliptical isothermal $\beta$-models obtaining for the best 
fit parameters $\beta^{circ}=0.728^{+0.025}_{-0.022}$, 
$\sigma_c^{circ}=0.679^{+0.045}_{-0.039}$ arcmin and 
$\beta^{ell}=0.737^{+0.027}_{-0.022}$, $\sigma_c^{ell}=0.746^{+0.044}_{-0.044}$ arcmin ($\sigma_c^{ell}$ is the core radius along the major axis), respectively, with the latter model providing a 
considerably better fit.

\begin{figure}[ht]
  \begin{center}
    \epsfig{file=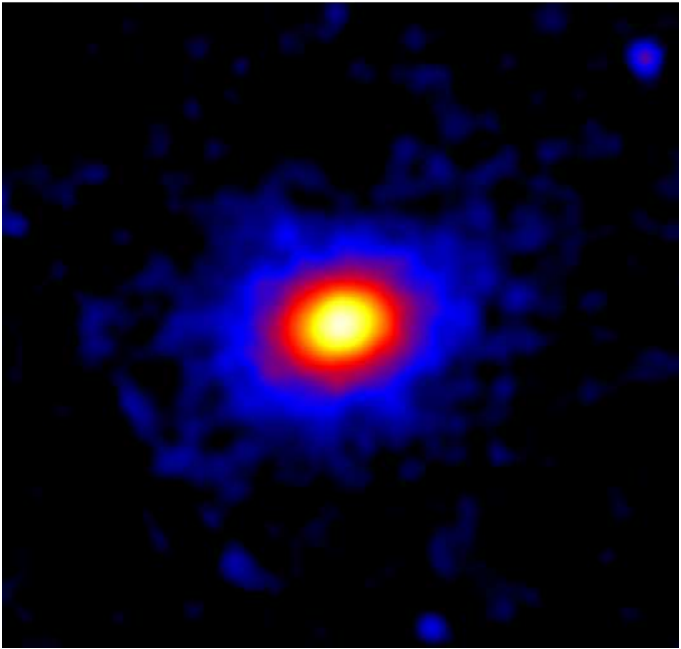, width=9cm}
  \end{center}
\caption{Chandra image of the cluster RBS797. The cluster is 
rather regular with, however, an ellipticity of 1.3-1.4 in E-W direction. The 
center and the position angle ($\approx -70^{\circ}$, N over E) of the 
various isophotes are almost the same over the entire radius range, from 
Schindler et al. (2001). } 
\end{figure}

More recently, \cite{schin2001} and \cite{defili2002} 
with a \textit{Chandra} observation of 
the galaxy cluster RBS797 revealed a pronounced aspherical geometry.
The analysis of the image (see Figure 1) gives a strong ellipticity, where 
the axis ratio of major to minor axis varies slightly from 1.3 at a radius of 
0.26 arcmin to 1.4 at a radius of 1.7 arcmin. A preliminary 
analysis of the surface brightness profile for 
RBS797 gives best fit parameters: $\beta^{circ}=0.62^{+0.03}_{-0.03}$, 
$\sigma_c^{circ}=7.32^{+0.7}_{-0.7}$ arcsec, 
and $\beta^{ell}=0.59^{+0.02}_{-0.02}$, $\sigma_c^{ell}=7.89^{+0.9}_{-0.9}$ 
arcsec (along the major axis), for the circular and elliptical models 
respectively.
\\
Assuming hydrostatic equilibrium and isothermal profile, the total mass of a 
cluster can be estimated from the parameters provided by the surface 
brightness fit and clearly some care is needed if 
the galaxy cluster in question shows a 
pronounced ellipsoidal shape. In this case, 
the general expression for the total mass density $\rho_{tot}$ is given by
\begin{equation}
\rho_{tot}=-\Big(\frac{k T_{gas}}{4\pi G \mu m_p}\Big)\, \Delta 
\Big[ \, {\rm{ln}}  \rho_{gas} \, \Big],
\end{equation}
where $G$ is the gravitational constant, $k$ the Boltzman constant, 
$\rho_{gas}$ the gas mass density, $T_{gas}$ its 
temperature and $\mu m_p$ is the mean particle mass of the gas.\\
In order to obtain the total mass of the cluster, one assumes a spherical 
geometry and integrates this equation over a sphere with radius $R$. 
\\
For RBS797 we have additionally investigated the ellipsoidal geometry. 
For the unknown extent in the line-of-sight we assume either the visible 
minor or major axis (i.e. oblate or prolate shape) and integrated the equation over an ellipsoid (concentric and with the same axis ratios) with major 
semi-axis $R$. We thus obtained mass estimates for the spherical shape: 
$$
M_{tot}^{sph}(R=4 \sigma_c^{circ}=29.28 \ {\rm arcsec})=8.66^{+2.5}_{-2.3} 
\times 10^{13} \ {\rm M}_{\odot}
$$ 
and 
$$
M_{tot}^{sph}(R=30 \sigma_c^{circ}=219.6 \ {\rm arcsec})=6.89^{+2.0}_{-1.8} 
\times 10^{14} \ {\rm M}_{\odot}
$$ 
and for ellipsoidal shapes, which are, compared at the same values of $R$, lower than those for the spherical symmetry by $\sim 10 \%$ and $\sim 17 \%$ for oblate or prolate shapes, respectively.  
For a detailed analysis of the influence of the ellipsoidal shape of the 
intracluster medium distribution on the total mass and gas mass fraction 
estimations see \cite{pif2002}.

\section{SZ effect and Hubble constant}
The SZ effect is difficult to measure, since systematic errors 
can be important.  For example, \cite{ina1995} analysed the reliability of 
the Hubble constant measurement based on the SZ effect. \cite{coo1998} showed 
that projection effects of clusters can lead incidence on the calculations of 
the Hubble constant and the gas mass fraction, and \cite{Hu1998} as well as 
\cite{su1999} pointed out, that galaxy cluster shapes can produce systematic 
errors on the measured value of $H_o$.
\\
We assume an ellipsoidal $\beta$-model
\footnote{The set of coordinates $r_x$, $r_y$ and $r_z$, as well as the 
characteristic lengths of the half axes of the ellipsoid $\zeta_1$, $\zeta_2$ 
and $\zeta_3$ are defined
in units of the core radius $r_c$.}:
\begin{equation}
n_e(r_x, r_y, r_z) = 
n_{eo} \left[ 1 + \frac{r_x^2}{\zeta_1^2} + \frac{r_y^2}{\zeta_2^2}
+ \frac{r_z^2}{\zeta_3^2} \right ]^{-3 \beta /2 }~,\label{eq:dp}
\end{equation}
where $n_{eo}$ is the electron number density at the center of the cluster and 
$\beta$ is a free fitting parameter which lies in the range $1/2 
\leq \beta \leq 1$.
\\
The Compton parameter $y$ and the X-ray surface brightness $S_X$ depend on 
the temperature of the hot gas $T_e$ and the electron number density $n_e$ 
as follows  
\begin{equation}
y  \propto  2  \, \int_{0}^{l} \, n_e T_e dr_y~,
\end{equation}
\begin{equation}
S_x  \propto  2  \, \int_{0}^{l} \, n_e^2 \, \sqrt{T_e} dr_y~,
\end{equation}
where $l$ is the maximal extension of the hot gas along the line of sight 
in units of the core radius $r_c$. 
We have chosen the line of sight along the $r_y$ axis.\\
For a detailed calculation of the Compton parameter and the X-ray surface 
brightness with an isothermal profile $T_e=T_{eo}$, 
we refer to the paper by \cite{Puy2000}:  
\begin{eqnarray}
y(r_x,r_z) &\equiv& 
T_{eo} n_{eo} \zeta_2 r_c
 \times   \left( 1+\frac{r_x^2}{\zeta_1^2} + \frac{r_z^2}{\zeta_3^2} 
\right)^{-\frac{3}{2}\beta + \frac{1}{2}} \nonumber \\
&\times& \, {\Gamma}_y(\beta,m)
\end{eqnarray}
and 
\begin{equation}
{\Gamma}_y(\beta,m) \, = \, 
\left[ {\cal{B}}\left(\frac{3}{2}\beta-\frac{1}{2},\frac{1}{2}\right) -
{\cal{B}}_m\left(\frac{3}{2}\beta-
\frac{1}{2},
\frac{1}{2}\right)\right], \nonumber
\end{equation}
\begin{eqnarray}
S_X(r_x,r_z)&\equiv& 
\frac{n_{eo}^2 \sqrt{T_{eo}} \zeta_2 r_c}{(1+z)^3} \times 
\left( 1+\frac{r_x^2}{\zeta_1^2} + \frac{r_z^2}{\zeta_3^2} 
\right)^{-3\beta + \frac{1}{2}} \nonumber \\
&\times& \,
{\Gamma}_{S_X}(\beta,m) \nonumber
\end{eqnarray}
at the redshift $z$, with
\begin{equation}
{\Gamma}_{S_X}(\beta,m) \, = \, 
\left[ {\cal{B}}\left(3\beta-\frac{1}{2},\frac{1}{2}\right) 
-{\cal{B}}_m \left(3\beta-\frac{1}{2} ,
\frac{1}{2}\right)\right] \nonumber
\end{equation}
where we introduced the Beta $\cal{B}$ and the incomplete Beta-functions 
${\cal{B}}_m$ 
with the cut-off parameter $m$ given by:
\begin{equation}
m =  \frac{1+(r_x/\zeta_1)^2 + (r_z/\zeta_3)^2}
{1+(r_x/\zeta_1)^2 + (r_z/\zeta_3)^2 + (l/\zeta_2)^2}~.
\end{equation} 
Introducing the angular core radius $\theta_c=r_c/D_A$, where $D_A$ is 
the angular diameter distance of the cluster, 
we can estimate the Hubble constant from the ratio between $y^2(r_x,r_z)$ 
and $S_X (r_x,r_z)$.
If we choose the line of sight through the cluster center we get, 
see \cite{Puy2000}:
\begin{equation}
H_0(l) \, \equiv \,  T_{eo}^{3/2} \, \alpha(l) \, \theta_c \, 
\frac{\Bigl[{\Gamma}_{y}(\beta,m)\Bigr]^2}
{{\Gamma}_{S_X}(\beta,m)}
\end{equation}
with $\alpha(l) =S_X(l)/y^2(l)$ 
for a finite extension $l$ and, for an infinitely extended cluster, we get instead  
\begin{equation}
H_0 (\infty) \, \equiv \,   T_{eo}^{3/2} \, \alpha(\infty)
\, \theta_c \, 
\frac{\Bigl[{\Gamma}_{y}(\beta,0)\Bigr]^2}
{{\Gamma}_{S_X}(\beta,0)}
\end{equation}
with $\alpha(\infty)= S_X(\infty)/y^2(\infty)$.
 Since $S_X$ and $y^2$ are observed quantities, the ratios 
$\alpha(\infty)$ and $\alpha(l)$ are in the following both set equal 
to the measured value $\alpha_{obs}$.
\\
\cite{maus2000} determined $H_o$ from $X$-ray measurements 
of A1835 obtained with ROSAT and from the corresponding millimetre 
observations of the SZ effect with the {\it Suzie} experiment. Assuming an 
infinitely extended, spherical gas distribution with an isothermal profile 
$\beta=0.58 \pm 0.02$, $T_{eo}=9.8 ^{+2.3}_{-1.3}$ keV, 
$n_{eo} = 5.64 ^{+1.61}_{-1.02} \times 10^{-2}$ cm$^{-3}$, they 
found $H_o=59^{+36}_{-28}$ km s$^{-1}$ Mpc$^{-1}$. The figures 
on the right show the influence of geometry and of the assumption of finite 
extension on the above result using the same input parameters. The left 
figure 2 shows that for a spherical geometry $H_o$ displays a strong 
dependence on the cluster extension. The right figure 2 gives the value of 
$H_o$ assuming an infinite extended ellipsoid shaped cluster as a 
function of its axis ratio $\zeta_1/\zeta_3$. We find $H_o \sim 51.6$ 
km Mpc$^{-1}$ s$^{-1}$ for $\zeta_1/\zeta_2 = 1.5$.
\\
Although the isothermal distribution is often a reasonable approximation of 
the actual observed clusters, some clusters do show non-isothermal features. 
A polytropic profile has the following 
form:
\begin{equation}
T_e \, = \, T_{eo} \, \Bigl[ \frac{n_e}{n_{eo}} \Bigr]^{\gamma-1}.
\end{equation}
The isothermal profile is obtained by setting $\gamma=1$.
The polytropic profile can play a role in the estimates of the Hubble constant.
 We refer to the paper of \cite{Puy2000}. The figure 3 shows that a small 
deviation from the isothermal case ($\gamma=1$) leads to a significant change 
in the values of the Hubble constant. For instance, assuming an isothermal 
instead of a polytropic profile with index $\gamma=1.2$ leads to an 
overestimation of 34 \% for $H_o$. Indeed for A3562 a polytropic profile 
with index $\gamma =1.16 \pm 0.03$ has been determined by 
\cite{ett2000}.
\vskip2mm

\begin{figure}[ht]
  \begin{center}
    \epsfig{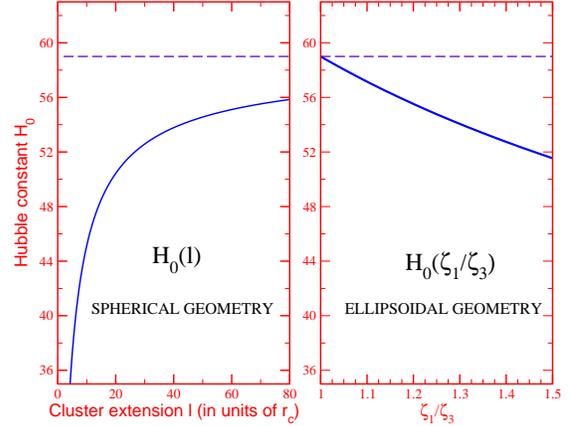}
  \end{center}
\caption{The dotted line corresponds to the value of 
the Hubble constant $H_o=59$ km s$^{-1}$ Mpc$^{-1}$ derived from the data of 
Mauskopf et al. (2000). 
The left figure shows the influence of finite extension, while the 
right figure gives the value of $H_o$ assuming an axisymmetric ellipsoidal 
geometry. In the latter case, oblate or prolate geometry give the same value 
of $H_o$ when taking a line of sight through the cluster center, as is 
assumed here.}  
\end{figure}

\begin{figure}[ht]
  \begin{center}
    \epsfig{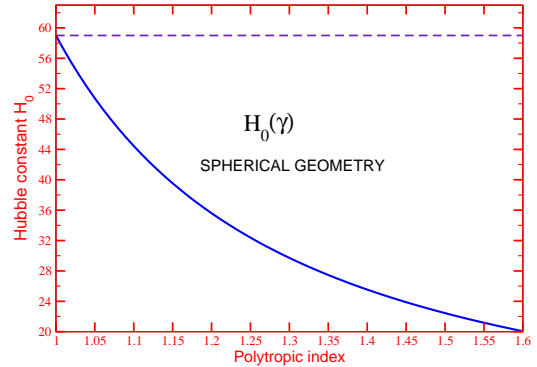}
  \end{center}
\caption{The Hubble constant (in km s$^{-1}$ Mpc$^{-1}$) 
for different values of the polytropic index. The 
line of sight is taken to go through the center of the cluster, which is 
assumed to have a spherical profile with {\it infinite} extension 
($\beta=2/3$-model). $H_o=59$ km s$^{-1}$ Mpc$^{-1}$ is the dotted line 
according to Mauskopf et al. (2001).}  
\end{figure}

\section*{Outlook}
We see thus that the geometry of galaxy clusters can affect 
the estimates of the masses and can substantially change the analysis of the 
SZ effect. The temperature profile is another important factor of change, indeed 
the hypothesis of isothermal profile might be too simple for some 
clusters. 
\\
The recent XMM observation of the Coma cluster by 
\cite{arn2001} points out that the temperature distribution is 
remarkably homogeneous suggesting that the core is actually in a 
relaxed state, but also that the NGC 4839 group is falling into the Coma 
cluster, see \cite{neu2001}. This last result suggests the presence of more complex
substructures which can be explained by a first infall onto the Coma cluster. 
\\
In some clusters there is evidence of cooling flows in the central 
regions. 
Cooling flows in galaxy clusters can substantially change the 
temperature profiles, especially in the inner regions. \cite{kei1991} 
and \cite{maju2000} 
investigated the changes induced by a cooling flow in the 
temperature and density profiles, and their implication on the SZ effect. 
An analogous effect might be observed if the fluid speed and density are 
higher, as it is expected for merging processes in clusters of galaxies. 
For example very recently \cite{defili2002} revealed that the CL 
0939+4713 cluster has an irregular morphology with evident substructures 
which seem to be in the process of merging.  
\\
The existence of a magnetic field in the intra-cluster medium might 
also be important. A first analysis by \cite{ko2002} suggests a possible 
influence on the SZ effect. 
\\
In these contexts an aspherical distribution of the density could play 
an additional important role. In summary, we see that it is crucial to know 
the shape of a cluster 
and its temperature profile. To address these problems, the Chandra and XMM-Newton 
satellites have the necessary spatial resolution for a complete 
analysis of the geometrical parameters of galaxy clusters.  

\begin{acknowledgements}

We would like to thank Fred Jansen for organizing this enjoyable and 
stimulating meeting. This work has been supported by the 
{\it Tomalla-Stiftung} and by the Swiss National 
Science Foundation.

\end{acknowledgements}


\begin{thebibliography}{}

\bibitem[\protect\astroncite{Allen et al.}{(1993)}]{Al1993}
Allen, S., Fabian, A., Johnstone, D, White, D., Daines, S., 
Edge, A., Steward, G., 1993, MNRAS 262, 901

\bibitem[\protect\astroncite{Arnaud et al.}{(2001)}]{arn2001}
Arnaud, M., Aghanim, N., Gastaud, R., Neumann, D., Lumb, D., Briel, U., 
Altieri, B., Ghizzardi, S., Mittaz, J., Sasseen, T., Vestrand, W., 2001, 
A\&A 365, L67

\bibitem[\protect\astroncite{Cavaliere \& Fusco-Femiano}{(1976)}]{cava1976}
Cavaliere, A., Fusco-Femiano, R., 1976, A\&A 49, 137

\bibitem[\protect\astroncite{Cooray}{(1998)}]{coo1998}
Cooray, A., 1998, A\&A 339, 623

\bibitem[\protect\astroncite{De Filippis, Schindler and Castillo-Morales}
{(2002)}]{defili2002}
De Filippis E., Schindler S., Castillo-Morales A., 2002, 
ESA Proceedings Symposium {\it New visions of the X-Ray Universe in the 
XMM-Newton and Chandra era}, ESA SP-488, astro-ph/0201349

\bibitem[\protect\astroncite{Ettori et al.}{(2000)}]{ett2000}
Ettori, S., Bardelli, S., De Grandi, S., Molendi, S., Zamorani, G., 
Zucca, E., 2000, MNRAS 318, 239

\bibitem[\protect\astroncite{Fabricant, Rybicki and Gorenstein}{(1984)}]
{Fab1984}
Fabricant, D., Rybicki, G., Gorenstein P., 1984, ApJ 286, 186 

\bibitem[\protect\astroncite{Hughes \& Birkinshaw}{(1998)}]{Hu1998}
Hughes, J., Birkinshaw M., 1998, ApJ 501, 1 

\bibitem[\protect\astroncite{Inagaki, Suginohara and Suto}{(1995)}]{ina1995}
Inagaki, Y., Suginohara, T., Suto, Y., 1995, Publ. Astron. Soc. Japan 47, 411

\bibitem[\protect\astroncite{Koch, Jetzer and Puy}{(2002)}]{ko2002}
Koch, P.M., Jetzer, Ph., Puy, D., 2002 in preparation

\bibitem[\protect\astroncite{Majumdar \& Nath}{(2000)}]{maju2000}
Majumdar, S., Nath, B., 2000, MNRAS 324, 537

\bibitem[\protect\astroncite{Mauskopf et al.}{(2000)}]{maus2000}
Mauskopf, P., Ade, P., Allen, W., Church, S., Edge, A., Ganga, K., 
Holzapfel, W., Lange, A., Rownd, B., Philhour, B., Runyan, M., 2000, 
ApJ 538, 505

\bibitem[\protect\astroncite{Neumann \& B\"ohringer}{(1997)}]{Neu1997}
Neumann, D., B\"ohringer, H., 1997,  MNRAS 289, 123

\bibitem[\protect\astroncite{Neumann et al.}{(2001)}]{neu2001}
Neumann, D., Arnaud, M., Gastaud, R., Aghanim, N., Lumb, D., Briel, U., 
Vestrand, W., Steward, G., Molendi, S., Mittaz, J., 2001 A\&A 365, L74

\bibitem[\protect\astroncite{Piffaretti, Jetzer and Schindler}{(2002)}]
{pif2002}
Piffaretti, R., Jetzer, Ph., Schindler, S., 2002 in preparation

\bibitem[\protect\astroncite{Puy et al.}{(2000)}]{Puy2000}
Puy, D., Grenacher, L., Jetzer, Ph., Signore, M., 2000, A\&A, 
363, 415

\bibitem[\protect\astroncite{Schindler}{(2001)}]{sabi2001}
Schindler, S., Proceedings of the Vulcano workshop 2001, 
astro-ph/0109040

\bibitem[\protect\astroncite{Schindler et al.}{(2001)}]{schin2001}
Schindler, S., Castillo-Morales, A., De Filippis, E., Schwope, A., 
Wambsganss, J., 2001, A\&A lett. 376, L27

\bibitem[\protect\astroncite{Schlickeiser}{(1991)}]{kei1991}
Schlickeiser, R., 1991, A\&A 248, L23

\bibitem[\protect\astroncite{Sulkanen}{(1999)}]{su1999}
Sulkanen, M., 1999, ApJ 522, 59

\bibitem[\protect\astroncite{Sunyaev \& Zel'dovich}{(1972)}]{zel1972}
Sunyaev, R., Zel'dovich, Y., 1972, Comments Astr. Space Phys. 4, 173

\end{thebibliography}
\end{document}